\documentclass[10pt,onecolumn,letterpaper]{article}

\usepackage{iccv}
\usepackage{epsfig}
\usepackage{graphicx}
\usepackage{amsmath}
\usepackage{amssymb}
\usepackage{booktabs}

\usepackage[breaklinks=true,bookmarks=false]{hyperref}

\iccvfinalcopy 

\def\iccvPaperID{****} 

\ificcvfinal\fi

\begin{document}

\title{Osteoporosis Prescreening using Panoramic Radiographs through a Deep Convolutional Neural Network with Attention Mechanism}

\author{Heng Fan$^{1}$, Jiaxiang Ren$^{2}$, Jie Yang$^{3}$, Yi-Xian Qin$^{4}$, and Haibin Ling$^{2}$\\
$^{1}$Department of Computer Science and Engineering, University of North Texas\\
$^{2}$Department of Computer Science, Stony Brook University \\
$^{3}$Division of Oral and Maxillofacial Radiology, Kornberg School of Dentistry, Temple University \\
$^{4}$Department of Biomedical Engineering, Stony Brook University 
}

\maketitle
\ificcvfinal\thispagestyle{empty}\fi

\begin{abstract}
   
   {\bf Objectives.} The aim of this study was to investigate whether a deep convolutional neural network (CNN) with an attention module can detect osteoporosis on panoramic radiographs.

   {\bf Study Design.} A dataset of 70 panoramic radiographs (PRs) from 70 different subjects of age between 49 to 60 was used, including 49 subjects with osteoporosis and 21 normal subjects. We utilized the leave-one-out cross-validation approach to generate 70 training and test splits. Specifically, for each split, one image was used for testing and the remaining 69 images were used for training. A deep convolutional neural network (CNN) using the Siamese architecture was implemented through a fine-tuning process to classify an PR image using patches extracted from eight representative trabecula bone areas (Figure~\ref{fig:dpr}). In order to automatically learn the importance of different PR patches, an attention module was integrated into the deep CNN. Three metrics, including osteoporosis accuracy (OPA), non-osteoporosis accuracy (NOPA) and overall accuracy (OA), were utilized for performance evaluation.
   
   {\bf Results.} The proposed baseline CNN approach achieved the OPA, NOPA and OA scores of 0.667, 0.878 and 0.814, respectively. With the help of the attention module, the OPA, NOPA and OA scores were further improved to 0.714, 0.939 and 0.871, respectively.
   
   {\bf Conclusions.} The proposed method obtained promising results using deep CNN with an attention module, which might be applied to osteoporosis prescreening.
   
\end{abstract}


\section{Introduction}

As one of the most common bone diseases, osteoporosis has affected millions people every year~\cite{wright2014recent}. It thins and weakens bones by decreasing bone mineral density~\cite{kanis1994diagnosis}, and thus easily increases the risk of bone fractures. In order to reduce such risk, many approaches have been proposed to provide an early diagnosis of osteoporosis. Among them, the widely adopted method is to directly measure the bone mineral density of a patient using dual-energy X-ray absorptiometry (DXA)~\cite{kroger1992bone}. The patient is diagnosed with osteoporosis if the value of bone mineral density is less than a threshold. Despite being accurate, this method is expensive and inefficient, which heavily limits its application in routine examination.

Previous studies have demonstrated the close relationship between dental data and bone quality~\cite{horner1998relationship,leite2010correlations,calciolari2015panoramic,mansour2013panoramic,govindraju2014radiomorphometric,LingYLMXY14dmfr,white2005clinical}. Thus motivated, there are attempts to leveraging panoramic radiograph (PR) for potential of osteoporosis prescreening. In comparison with DXA, PR is cheaper and more efficient, making it more practical in routine examination. The method of~\cite{kavitha2012diagnosis} proposed to use the support vector machine (SVM)~\cite{CortesV95ml} method for osteoporosis detection by estimating bone mineral density. The algorithm of~\cite{kavitha2013combination} combined a histogram-based clustering algorithm and SVM for the diagnosis of osteoporosis. The approach of~\cite{lee2019osteoporosis} analyzed the texture of mandibular cortical bone on PR for osteoporosis diagnosis. The approach of~\cite{cakur2008dental} employed a statistic method on PR images for osteoporosis analysis. 

Despite progresses made in the aforementioned methods, it remains challenging to directly diagnose the osteoporosis condition using a global PR image due to many noisy background regions in the image. Inspired by the close correlation between trabecular bone structure and osteoporosis~\cite{eriksen1986normal,li2014trabecular}, researchers proposed to leverage trabecular patches in PR images for osteoporosis analysis. The trabecular patches were extracted based on trabecular bone areas manually annotated in an PR image. The method of~\cite{bo2017osteoporosis} introduced a two-stage SVM approach to apply PR patches for osteoporosis classification. Despite  promising results, this method leveraged hand-crafted features~\cite{dalal2005histograms,ojala2002multiresolution} to represent each PR patch, which might be sensitive to the diverse appearances of PR images.

Recently, thanks to its power in feature representation, deep convolution neural network (CNN) has been widely applied to various vision tasks, such as image recognition~\cite{krizhevsky2012imagenet,szegedy2015going}, object detection~\cite{girshick2014rich,ren2015faster} and segmentation~\cite{long2015fully,ronneberger2015u}. Thus inspired, some researchers have proposed to leverage deep CNN for osteoporosis analysis. The algorithm of~\cite{lee2019osteoporosis} employed a deep convolutional neural network for osteoporosis detection from a PR image. However, this method might ignore more reliable PR patches by trabecula landmarks for classification. The approach of~\cite{chu2018using} improved the performance of~\cite{bo2017osteoporosis} by replacing hand-crafted features with more powerful deep learning features. The method of~\cite{yumultitask} applied multi-task learning to further boost the performance. Despite the improvement, this approach treated each PR patch equally, which might ignore the fact that each patch plays a different role for the classification of different input.

Addressing the above problem, we introduced an attention module~\cite{VaswaniSPUJGKP17nips,RushCW15,bahdanau2014neural} into a deep convolutional neural network for osteoporosis prescreening. The objective of this investigation was to evaluate a  adaptive deep learning-based approach that was able to automatically assign weights to different PR patches for better osteoporosis analysis. We hypothesized that the data-driven deep CNN with attention module could learn to assign more importance to useful and discriminative PR patches while less weight to irrelevant ones, leading to more accurate results.

It is worth noting that, although deep learning has been applied to osteoporosis analysis in previous studies~\cite{chu2018using,yumultitask}, our study is significantly different. In particular, we proposed to utilize an attention module to adaptively assign weight to different PR patches for better accuracy. To the best of our knowledge, such attention mechanism has never be applied to image-based osteoporosis analysis. Moreover, the original dataset used in~\cite{chu2018using,yumultitask} contains 108 PR images, while in this paper a refined dataset of 70 PR images is used in the experiments.

\section{Materials and Methods}

\begin{figure*}
	\centering
	\includegraphics[width=0.8\linewidth]{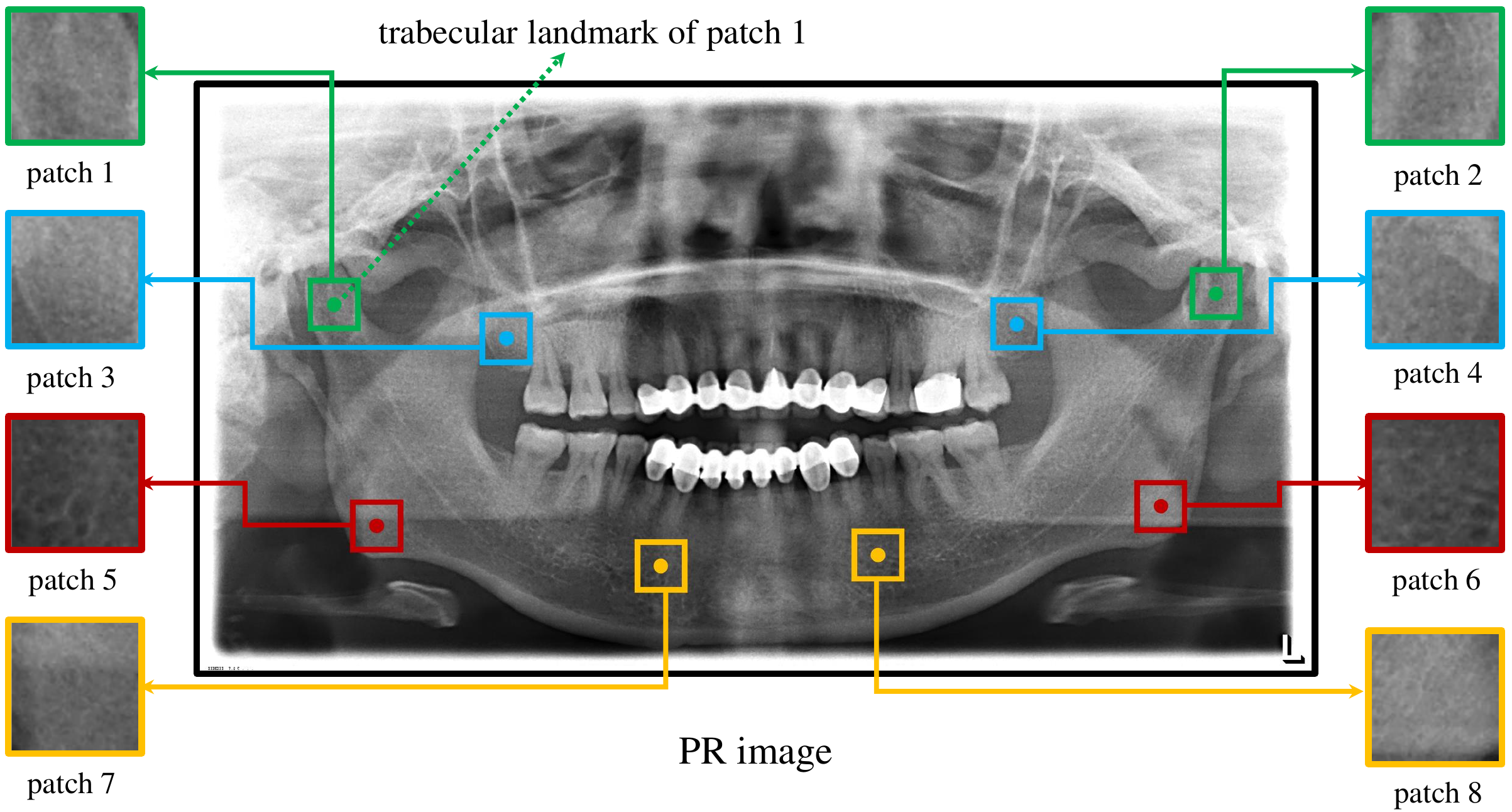}
	\caption{Illustration of region of interest (ROI) patch extraction in an PR image. The eight representative trabecular bone areas include the mandibular condyle, angle of the mandible, mandibular premolar region, and maxillary tuberosities.}
	\label{fig:dpr}
\end{figure*}

\subsection{Data preparation}

The PR images used in this study were from~\cite{chu2018using}, and no patient ID was involved. Consequently, this study is retroactive and no IRB is needed. The dataset contains 70 PR images from 70 different subjects (age 49-60 years; median age 53 years) for this study. All 70 PR images were collected from previous dental visits. Each subject was classified as normal (T-score $\ge$ -1.0) or osteoporosis (T-score $\le$ -2.5) according to the World Health Organization criteria~\cite{kanis1994assessment}. The PR image of each subject was obtained by an Orthophos XG\_5 machine (Dentsply Sirona, USA). In final, we compiled a dataset that consisted of 49 PR images labeled as osteoporosis and 21 PR images labeled as normal (\ie, without osteoporosis). Each PR image was manually annotated with eight representative trabecula bone areas (landmarks), four on each side, by an oral radiologist, and each landmark was represented with two coordinates on the PR image. Based on these landmarks, we extracted eight regions of interest (ROIs) to represent an PR image (see Figure~\ref{fig:dpr}).

\begin{figure}
	\centering
	\includegraphics[width=0.6\linewidth]{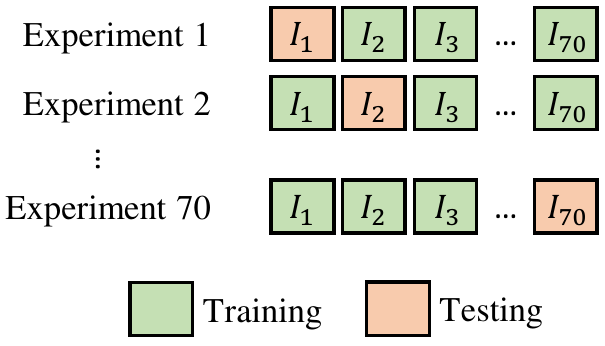}
	\caption{Illustration of leave-one-out cross-validation strategy for experiments. The $I_{i}$ represents the $i^{\mathrm{th}}$ ($1 \le i \le 70$) PR image.}
	\label{fig:training}
\end{figure}

We applied the leave-one-out cross-validation strategy to generate training and testing split for experiments in this study. In specific, in each experiment, 69 PR images were used for training and the rest one PR image for testing. This way, we conducted 70 experiments in total, as illustrated in Figure~\ref{fig:training}. 

\subsection{Data processing and augmentation}

From each PR image, eight ROI patches were extracted centered at the trabecula landmarks with a size of 100$\times$100 pixels (see Figure~\ref{fig:dpr}). In order to adapt to network input, we resized each PR patch to a size of 224$\times$224 pixels. To reduce overfitting, we utilized data augmentation including {\it random stretch} and {\it image flipping} in training. It is worth noting that, since each PR image was labeled by an oral radiologist, the trabecular landmarks in the PR image were guaranteed to be well calibrated for reliable experimental analysis.

\begin{figure*}
	\centering
	\includegraphics[width=0.8\linewidth]{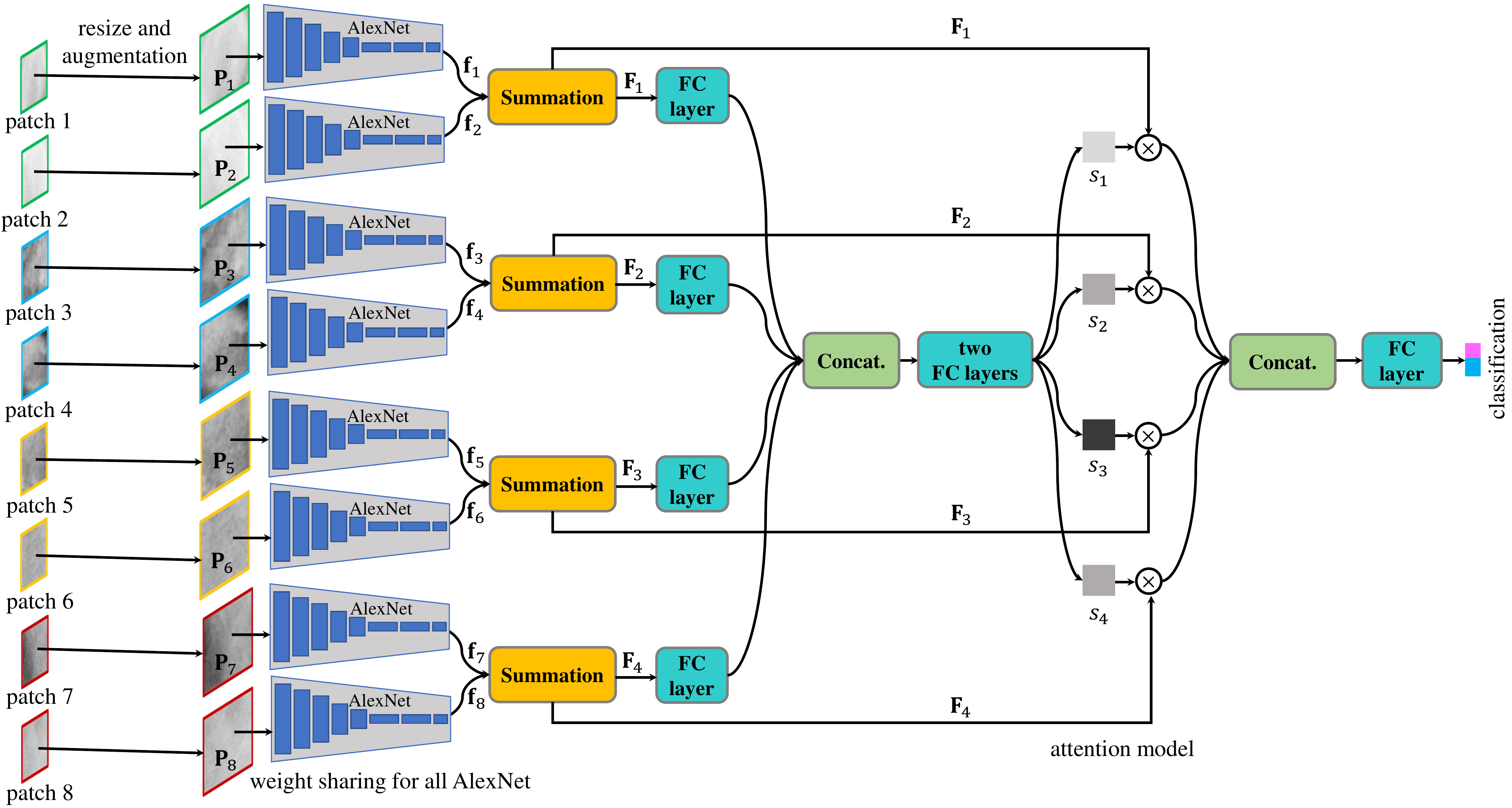}
	\caption{Illustration of the proposed method. $\mathrm{\bf P}_{i}$ represents the $i^{\mathrm{th}}$ ($1\le i \le 8$) processed PR patch, and ${\bf f}_{i}$ denotes the feature for $\mathrm{\bf P}_{i}$. $\mathrm{\bf F}_{i}$ represents the fused feature for group $i$, and $s_i$ denotes the learned weight for $\mathrm{\bf F}_{i}$. ``FC'' and ``Concat.'' mean \textit{fully connected} layer and \textit{concatenation} operation, respectively.}
	\label{fig:arch}
\end{figure*}

\subsection{Deep CNN with Attention Module for Osteoporosis Classification}

We formulated osteoporosis prescreening problem, \ie, predicting an PR image as either {\it osteoporosis} or {\it normal}, as a binary classification problem. Instead of using the entire PR image, we explored eight PR patches extracted based on trabecula landmarks for classification. For this purpose, we proposed a deep convolutional neural network with an attention module to perform classification. Figure~\ref{fig:arch} demonstrates the overall architecture of our method. 

In specific, we first resized each PR patch image from the size of 100$\times$100 pixels to 224$\times$224 pixels to adapt to network input. Then augmentation techniques were applied to obtain the network inputs. Assume that $\mathrm{\bf P}_{i}$ represents the $i^{\mathrm{th}}$ ($1\le i \le 8$) processed PR patch. Considering that each patch $\mathrm{\bf P}_{i}$ is cropped from the same PR image and thus shares similar low-level feature representation, we introduced a Siamese-style framework~\cite{Chicco20} that employed eight identical subnet branches for feature extraction of eight PR patches. Each subnet was implemented by the AlexNet~\cite{krizhevsky2012imagenet}. It is worth noting that, other network architectures (\eg, \cite{szegedy2015going,he2016deep}) can be used to replace AlexNet as well. In short, the feature extraction for $\mathrm{\bf P}_{i}$ can be written as
\begin{equation}
	{\bf f}_{i} = \phi_{\mathrm{Alex}}(\mathrm{\bf P}_{i}, \mathrm{\bf W}_{\mathrm{Alex}}),
\end{equation}
where ${\bf f}_{i}$ denotes the feature for $\mathrm{\bf P}_{i}$ and $\mathrm{\bf W}_{\mathrm{Alex}}$ represents the parameters of AlexNet. 

Because of the symmetry structure (see Figure~\ref{fig:dpr}), the eight patches of an PR image can be divided four groups, \ie, group $G_1 = \{\mathrm{\bf P}_{1},\mathrm{\bf P}_{2}$\}, group $G_2 = \{\mathrm{\bf P}_{3},\mathrm{\bf P}_{4}$\}, group $G_3 = \{\mathrm{\bf P}_{5},\mathrm{\bf P}_{6}$\} and group $G_4 = \{\mathrm{\bf P}_{7},\mathrm{\bf P}_{8}$\}. In each group, the two patches have similar features due to their similar appearances. Taking this into account, we fused the features for patches in the same group by summation. In specific, for $G_{1}$, we can obtain the fused feature $\mathrm{\bf F}_{1}$ of $\mathrm{\bf P}_{1}$ and $\mathrm{\bf P}_{2}$ via
\begin{equation}
	\mathrm{\bf F}_{1} = {\bf f}_{1} + {\bf f}_{2}.
\end{equation}
Likewise, we can compute the fused features $\mathrm{\bf F}_{2}$, $\mathrm{\bf F}_{3}$ and $\mathrm{\bf F}_{4}$ for patches in $G_2$, $G_3$ and $G_4$, respectively.

Instead of treating equally the fused feature of each group, we argued that, for different subject, each patch may play a different role in classifying an PR image. In order to effectively utilize the feature of each group, we introduced an attention module~\cite{VaswaniSPUJGKP17nips,RushCW15,bahdanau2014neural} which learns to automatically assign importance to PR patches in different groups. To this end, we first utilized a fully connected (fc) layer to transform the fused features of patches, and then concatenated the transformed features. The concatenated feature was fed to two consecutive fc layers to obtain a 4-dimension attention vector ${\bf S}=[s_{1}, s_{2}, s_{3}, s_{4}]$, which $s_{i}$ represented the learned weight for PR patches in group $G_{i}$. With $s_{i}$, we were able to obtain an adaptive feature ${\bf \Phi}$ by concatenating $\mathrm{\bf F}_{i}$ based on its weight via
\begin{equation}
	{\bf \Phi} = \mathrm{cat}(s_{1}\cdot\mathrm{\bf F}_{1}, s_{2}\cdot\mathrm{\bf F}_{2}, s_{3}\cdot\mathrm{\bf F}_{3}, s_{4}\cdot\mathrm{\bf F}_{4}),
\end{equation} 
where $\mathrm{cat}()$ represents the concatenation function. 

Finally, we employed ${\bf \Phi}$ for the classification of corresponding PR image. Specifically, we applied a fc layer on ${\bf \Phi}$ and then utilized the softmax function to classify ${\bf \Phi}$ to output the classification result $\tilde{\mathrm{\bf y}} \in \mathbb{R}^{1\times2}$. This classification process can be mathematically formulated as follows
\begin{equation}
	\tilde{\mathrm{\bf y}} = \mathrm{softmax}(\psi_{\mathrm{fc}}({\bf \Phi}, \mathrm{\bf W}_{\mathrm{fc}})),
\end{equation}
where $\mathrm{\bf W}_{\mathrm{fc}}$ denotes the parameters of fc layer. Notice that, the classification result $\tilde{\mathrm{\bf y}}=[\tilde{y}_\mathrm{op}, \tilde{y}_\mathrm{nop}]$ contains the probabilities $\tilde{y}_\mathrm{op}$ and $\tilde{y}_\mathrm{nop}$ of osteoporosis and non-osteoporosis. The final prediction for one subject was determined by the larger one between $\tilde{y}_\mathrm{op}$ and $\tilde{y}_\mathrm{nop}$.

Our network was implemented in PyTorch~\cite{paszke2019pytorch} on a computer with a single Nvidia GTX 1080 GPU with
8GB memory. We utilized cross entropy loss to train the network in an end-to-end manner using stochastic gradient descent (SGD). The initial parameters of AlexNet were borrowed from the pre-trained model on ImageNet~\cite{deng2009imagenet}. The learning rates for AlexNet and other layers were set to 0.0001 and 0.01 respectively. The momentum and weight decay are 0.9 and 0.0001. The batch size was 32. Our network was trained for 100 epochs in each experiment. The complete implementation of our approach will be released.

\subsection{Performance Evaluation Metric}

We utilize three metrics to measure the performance of our method, including osteoporosis accuracy (OPA), non-osteoporosis accuracy (NOPA) and overall accuracy (OA), which are defined as follows,

\begin{itemize}
	\item Osteoporosis accuracy (OPA) is defined as the ratio of correctly classified samples with osteoporosis to all osteoporosis samples.
	\item Non-Osteoporosis Accuracy (NOPA) is defined as the ratio of correctly classified non-osteoporosis samples to all non-osteoporosis samples.
	\item Overall accuracy (OA) is defined as the ratio of correctly classified samples to all samples. 
\end{itemize}

\subsection{Results}

\begin{table}[htbp]
	\centering
	\caption{Comparison of our algorithm with other classification approaches in osteoporosis analysis. The best result is highlighted in bold font.}
	\begin{tabular}{cccc}
		\toprule
		Method & OPA (\%)   & NOPA (\%)  & OA (\%) \\
		\midrule
		$k$-nearest-neighbor (KNN)~\cite{cover1967nearest}   & 65.3 & 42.9 & 58.6 \\
		Ensemble~\cite{opitz1999popular} & {\bf 79.6} & 19.1 & 61.4 \\
		Ours w/o attention & 66.7 & 87.8 & 81.4 \\
		Ours  & 71.4 & {\bf 93.9} & {\bf 87.1} \\
		\bottomrule
	\end{tabular}%
	\label{tab:addlabel}%
\end{table}%

Table~\ref{tab:addlabel} shows the performance of our method using OPA, NOPA and OA. In addition, we have also implemented other popular machine learning approaches for osteoporosis classification including $k$-nearest-neighbor (KNN)~\cite{cover1967nearest} and ensemble method~\cite{opitz1999popular}. We extracted LBP feature~\cite{ojala2002multiresolution} for PR patches in KNN and ensemble methods. The comparison of these algorithms are presented in Table~\ref{tab:addlabel}. From Table~\ref{tab:addlabel}, it could be observed that the classic classification models KNN and ensemble obtained promising results. In specific, the simple KNN method achieved OA score of 58.6\%, and the ensemble model obtained slightly better performance with 61.4\% OA score. Despite this, the results of NOPA in both KNN and ensemble methods were not accurate with 42.9\% and 19.1\% scores. In this study, we leveraged more powerful deep features for osteoporosis classification. Our baseline CNN method without attention module can significantly improve the OA score to 81.4\%, outperforming KNN and ensemble methods by 22.8\% and 20.0\%, respectively. Furthermore, to effectively combine different PR patches, an attention module was integrated into the deep network. The OPA, NOPA and OA scores have been boosted to 71.4\%, 93.9\% and 87.1\%, which clearly attested to the advantage of attention module.

\begin{figure}
	\centering
	\includegraphics[width=0.6\linewidth]{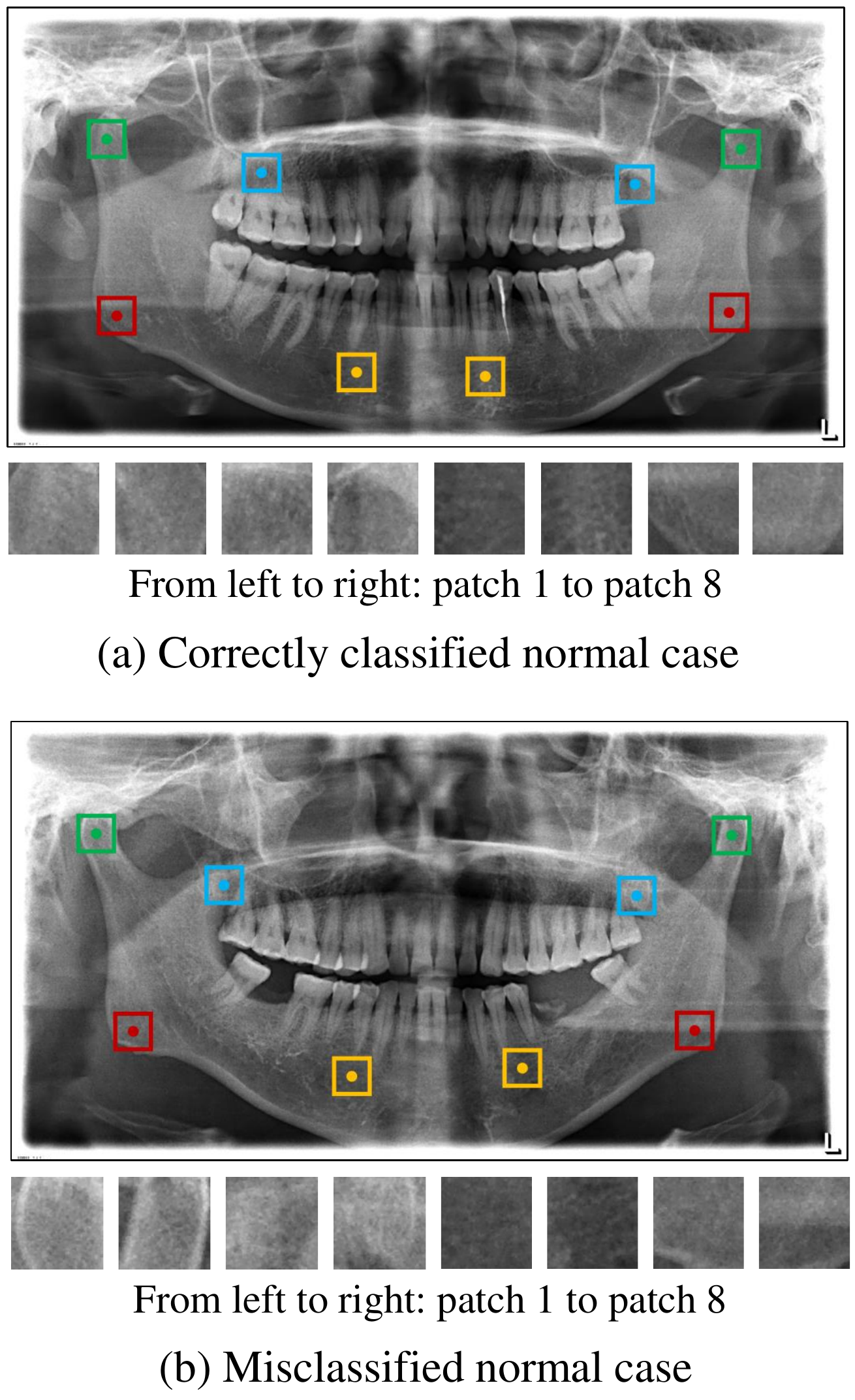}
	\caption{Visualization of selected results. The PR image in (a) is correctly classified normal case while the other in (b) is misclassified. The eight RoIs are shown below each PR image.}
	\label{fig:qual}
\end{figure}

\begin{figure}
	\centering
	\includegraphics[width=0.6\linewidth]{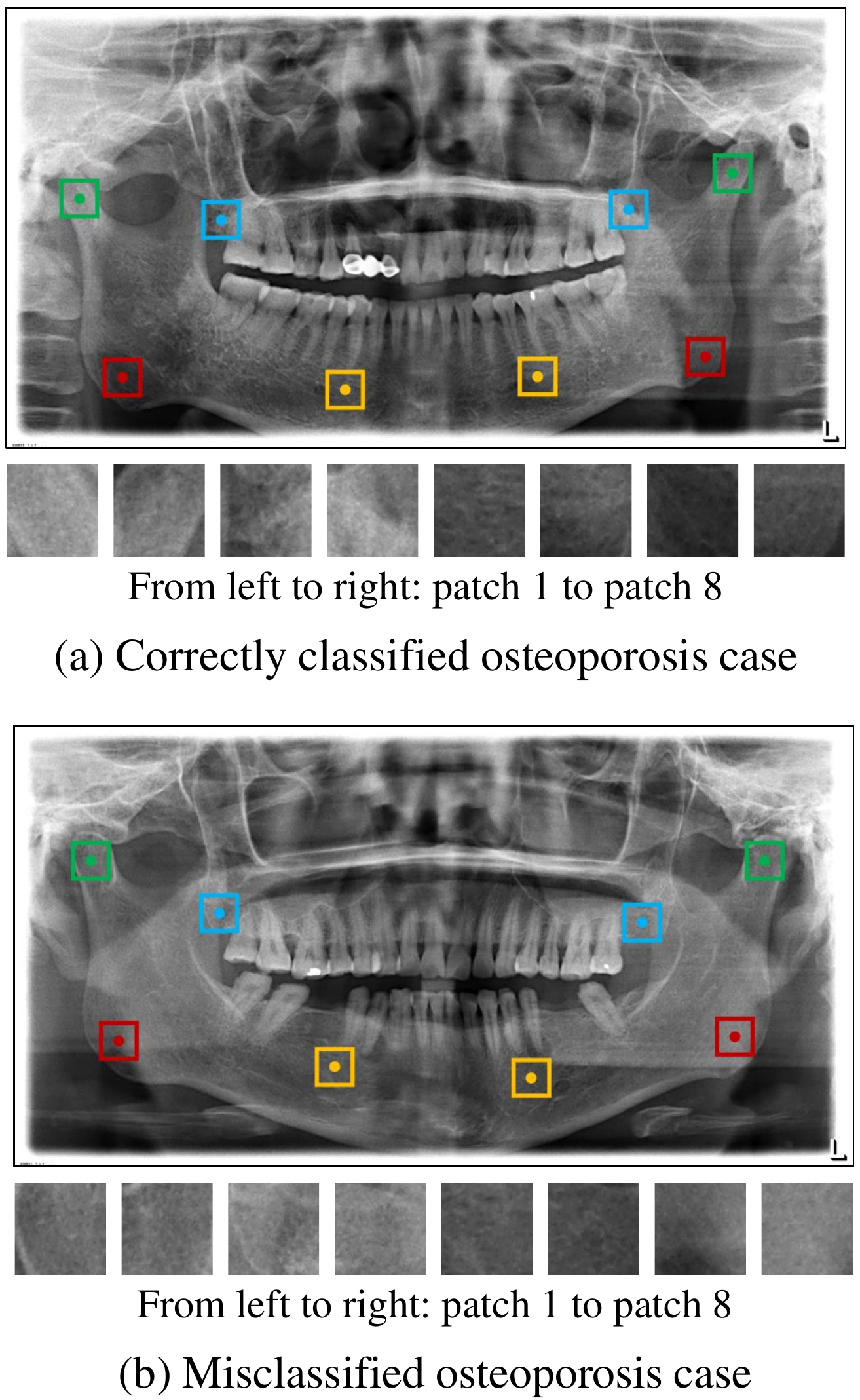}
	\caption{Visualization of selected results. The PR image in (a) is correctly classified osteoporosis case while the other in (b) is misclassified. The eight RoIs are shown below each PR image.}
	\label{fig:qual1}
\end{figure}

Figure~\ref{fig:qual} and~\ref{fig:qual1} show the qualitative results of the proposed method. In specific, Figure~\ref{fig:qual} (a) and (b) show the correctly classified and misclassified normal samples. Figure~\ref{fig:qual1} (a) and (b) show the correctly classified and misclassified osteoporosis samples. We observed that, the visual appearance of two failure cases in Figure~\ref{fig:qual} and~\ref{fig:qual1} were confusing, and it was prone to misclassify them.

\subsection{Discussion}

Osteoporosis is one of the most bone diseases. In recent decades, PR has drawn increasing interest in osteoporosis analysis. Compared with DXA measurement, PR is cheaper and more efficient, which makes it more suitable for a routine examination. Previous studies focused on analyzing the global PR image for osteoporosis~\cite{kavitha2012diagnosis,kavitha2013combination,lee2019osteoporosis,cakur2008dental}. However, these methods were difficult to reliably create a mapping from PR image to osteoporosis condition. 

Motivated by the close correlation between trabecular bone structure and osteoporosis~\cite{eriksen1986normal,li2014trabecular}, some researchers proposed to leverage trabecular landmarks to extract trabecular patches from PR images and applied a two-stage SVM method for osteoporosis~\cite{bo2017osteoporosis} analysis. Different hand-crafted features have been utilized for obtaining accurate performance. However, these hand-crafted features were sensitive to various appearances in PR images, which might affect classification results. To further improve the performance, deep learning method has been introduced to learn more powerful features for osteoporosis classification~\cite{chu2018using}. Despite promising results, this approach treated all PR patches with the same importance, which was not able to fully exploit each patch for classifying different PR images.

In this study, we proposed a deep network for osteoporosis classification with attention module. We utilized a Siamese architecture to simultaneously extract deep features for multiple PR patches. In order to effectively combine these patches for classification, we utilized the attention module to automatically learn the weights for different patches. Experimental results demonstrated that the attention module in our method significantly improved performance.

In conclusion, this study was designed to use PR images for osteoporosis prescreening. Specifically, we proposed a deep learning-based solution to explore PR patches extracted based on trabecula landmarks for osteoporosis classification. To effectively combine different PR patches, we developed an attention module to be integrated into the deep neural network for improvement. In the experiments, our method achieved the OA score of 87.1\%, outperforming non-deep-learning methods such as KNN~\cite{cover1967nearest} and ensemble~\cite{opitz1999popular}. This study showed that the proposed method had the potential to be applied in routine examination for inexpensive osteoporosis prescreening by using PR images.

\section*{Declaration of competing interest}

None declared.

{\small
\bibliographystyle{ieee_fullname}
\bibliography{egbib}
}

\end{document}